\documentstyle[12pt]{article}
\begin{document}
\newcommand{\ad}[1]{\mbox{${#1}^{\dagger}$}}
\newcommand{\kone}[1]{\mbox{$|{#1}\rangle$}}
\newcommand{\kttwo}[2]{\mbox{$|{#1},{#2}\rangle$}}
\newcommand{\kthree}[3]{\mbox{$|{#1}\,;\,{#2}_{#3}\rangle$}}
\newfont{\blackb}{msbm10 scaled 1200}
\def\BR{\mbox{\blackb R}}
\def\BC{\mbox{\blackb C}}
\title{{\bf Control of Wave Packet Revivals Using Geometric Phases}}
\author{{\bf S. Seshadri,}$^1$ {\bf S. Lakshmibala,}$^2$ {\bf and V.
Balakrishnan}$^2$ \\
{\em $^1$ The Institute of Mathematical Sciences, C.I.T. Campus,}\\ 
{\em Chennai
600 113, India.}\\ 
{\em $^2$ Department of Physics, Indian Institute of Technology - 
Madras,}\\
{\em Chennai 600 036, India.}}
\date{}
\maketitle
\begin{abstract}
Wave packets in a system governed by a Hamiltonian with a generic
nonlinear spectrum
typically exhibit both full and fractional revivals. It is
shown that the latter can be eliminated by inducing suitable geometric
phases in the states, by varying the parameters in the Hamiltonian
cyclically with a period $T$. Further, with the introduction of this
natural time step $T$, the occurrence of near revivals can be mapped onto
that of Poincar\'{e} recurrences in an irrational rotation map of the
circle. The distinctive recurrence time statistics of the
latter can thus serve as a clear signature of the dynamics of wave
packet revivals.
\end{abstract}
\mbox{}\hskip 1cm  near revivals; rotation map; recurrence
statistics. 
\newpage
\section*{\large 1. INTRODUCTION}
\renewcommand{\theequation}{1.\arabic{equation}}
\setcounter{equation}{0} 
\hskip .5cm There has been considerable interest in the dynamics of wave
packets right from the early days of quantum mechanics.$\,^{1)}$ In
recent years this subject has gained further impetus from
a variety of experiments in molecular
systems,$\,^{2)}$ Rydberg wave packets,$\,^{3)}$\,
semiconductor quantum wells,$\,^{4)}$\, etc., which require a detailed
understanding of wave
packet evolution -- e.g., investigations in molecular physics
(using femtosecond laser pulse techniques) based on the phenomena
of full and fractional revivals$\,^{5)}$\, of a vibrational wave packet.

Consider the evolution of an initial state \kone{\psi(0)} of a system
governed by a Hamiltonian $H$. In general, if \kone{\psi(0)} is not an
eigenstate of $H$, the correlation or overlap function
\begin{equation}
{\cal C}(t)\;=\; 
 |\langle \psi(0) \kone{\psi(t)}|^2
\end{equation}
will decrease from its initial value of unity as $t$ increases. Under
special circumstances, however, ${\cal C}(t)$ may return to its initial
value at some particular instant of time, and a revival is said to occur.
One might therefore expect that revivals would be facilitated if the
system is prepared in an initial state \kone{\psi(0)} which is a  
superposition (a ``wave packet") of stationary states of $H$, sharply
peaked about some energy eigenvalue $E_{n_0}$. (We consider, for
simplicity, the one-dimensional case.) In the {\em linear} case of an
equally-spaced spectrum, it is easy to show that revivals are a
consequence of a simple periodicity. In general, however, 
$E_{n}$ is a {\em nonlinear} function of $n$.
On expanding $E_n$ in a Taylor series about
$E_{n_0}$, it turns out that the revival times depend on the coefficients
of the
linear and quadratic terms in this expansion.$\,^{6)}$  Cubic and higher
order terms lead to  (rare)
``super-revivals"$\,^{7)}$\, and can be neglected in most realistic
situations, if the wave packet is peaked sufficiently sharply 
about $E_{n_0}$. The quadratic term in the Taylor expansion leads to
so-called {\em fractional}
revivals that could occur between two full revivals: 
the initial wave packet
evolves
to a state that can be described as a collection of a small number of 
spatially distributed
sub-packets, each of which closely reproduces the initial
state$\,^{8)}$. Experiments with NaI molecules display
complex wave packet evolution that suggests fractional revivals of
molecular wave packets.$\,^{9,10)}$ 

Revivals and fractional revivals of a wave packet  are, of course,
a manifestation of the interference between the  constituent basis states,
each of which acquires a different phase during its temporal evolution.
In general, the occurence of fractional revivals is contingent on delicate
arithmetic properties of the numerical values of the parameters that
appear in the Hamiltonian, e.g., the closeness of certain ratios of these
parameters to rational numbers. The question that arises then is whether
one can make the phenomenon of revivals more ``robust" by suppressing the
plethora of fractional or partial revivals in favour of clearly-signalled
full (or nearly full) revivals, even though the spectrum is nonlinear. We
shall show that this is indeed feasible, by exploiting the possibility of
inducing {\em geometric} (Berry) phases in the states
over and
above the dynamical ones acquired by unitary evolution. As is well known,
such phases may occur if at least two parameters in the Hamiltonian are
varied cyclically (with a period $T$, say) and adiabatically (the
original,
simplest setting for Berry phases$\,^{11)}$). This variation can be
tailored so as
to cancel out all fractional revivals in a generic nonlinear Hamiltonian.
The cycling of parameters, however, means that it is only at instants of
time separated by an interval $T$ that the Hamiltonian returns to its
original self. Therefore, with the introduction of this natural time step
$T$ into the problem, it is only meaningful to speak of possible revivals
of a state at the instants $T, \, 2T, \, \cdots$. It turns out that {\em
nearly} full revivals (``near revivals") are still possible after this
discretization of time.
Moreover, the statistical
distribution of these events is found to be precisely that of
Poincar\'{e} recurrences in the rotation map on a circle.

The plan of this paper is as follows: In the next section, we give a brief
review of wave packet revivals with particular emphasis on  fractional
revivals. In Section 3 we show that by inducing suitable
geometric phases in the basis states, we can eliminate all fractional
revivals. The formalism is also illustrated explicitly. Finally, in
Section 4 we
indicate briefly how the statistics of near revivals in the reduced
problem 
can be mapped onto that of
recurrences in an irrational rotation  of the circle. 

\section*{\large 2. FRACTIONAL REVIVALS: REVIEW} 
\renewcommand{\theequation}{2.\arabic{equation}}
\setcounter{equation}{0}
\hskip .5cm Consider a system with a time-independent hermitian 
Hamiltonian $H$, with spectrum $\{E_n\}$ and eigenstates
$\{\kone{\phi_n}\}$. Let the system be prepared in an initial state
\kone{\psi(0)} that is a
superposition of the $\{\kone{\phi_n}\}$, sharply peaked
about some $n_0$. As mentioned earlier, we 
expand $E_n$ as
\begin{equation}
E_n \;=\; E_{n_0} \, + (n-n_0)\, E'_{n_0} \, +  (1/2) (n-n_0)^2
\,E^{''}_{n_0}\,
+ \cdots \,
\end{equation}
As we wish to analyze only revivals and
fractional revivals,
we retain only terms up to  the second order in Eq. (2.1) and 
shift  $n$ by $n_0$ for notational simplicity, to arrive at the quadratic
form
\begin{equation}
E_n \; =\; C_0 \, + \, C_1 \, n \, + \, C_2 \, n^2\,.
\end{equation}
The coefficients $C_i$ 
evidently depend on  the parameters that occur in $H$. We shall assume
that $C_1,\, C_2 >\,0$. (The modifications necessary in other
cases are easily worked out.)
In the  \kone{\phi_n}-basis the time evolution operator
$U(t) = \exp \, [-i \, H \, t/\hbar]$ has the representation
\begin{equation}   
U(t) \;=\; \sum_n^{\infty} \; \exp  [ - i\,
( C_0 +C_1 \,n  + C_2\, n^2)t/ \hbar ] \;\; \kone{\phi_n} \, \langle   
\phi_n \, | \;\;.
\end{equation}
For a full revival (${\cal C}(t)\,=\,1$) to occur at time $t$,
 $U(t)$ must reduce  to the unit operator (apart from a possible  overall
phase factor), i.e.,  
$(C_1 n +C_2 n^2)t$ must be an integer multiple of $2 \pi \hbar$ for every
$n$
in the summation. The following cases arise: \\
(i) $C_1 \ne 0, \, C_2 =0$ (equi-spaced or linear
spectrum): Revivals of an initial state occur with a period 
$T_{rev} = 2 \pi \hbar/C_1$. \\
(ii) $C_1 = 0,\, C_2 \ne 0$: Revivals occur with a period
$T_{rev} = 2\pi\hbar /C_2$. \\
(iii) $ C_1, C_2 \ne 0$,  $C_1 / C_2\, =$  a rational
number $r/s$: Once again, full revivals occur with a fundamental revival
time  $T_{rev} = 2\pi \hbar s /C_2$. A specific example is provided by the
Hamiltonian ${\ad{a}}^2 a^2 \,=\, \ad{a}a (\ad{a}a -1)$ that is relevant
to wave packets  propagating in a Kerr
medium.$\,^{12)}$ It is evident that, in this case,
$E_n$ is proportional to $n(n-1)$ which is an even integer for every
$n$.\\
(iv) $ C_1, C_2 \ne 0$, $C_1 / C_2$ irrational (the generic case): As  the  
condition $(C_2 n^2 + C_1 n)t \,=\, 2 \pi \hbar m $ ($m$ = integer)
cannot be satisfied for {\em all} $n$  at any value of $t$,  full revivals
are no longer possible. However,  at certain instants of time the quantity
${\cal C}(t)$ could come {\em arbitrarily} close to unity, producing a
near revival. 

Between occurrences of  full (or near) revivals, the wave packet
 breaks up into a finite sum of subsidiary
packets at specific instants of time,  {\em provided} the
spectrum is  nonlinear, i.e., $C_2 \ne 0$.$\,^{8)}$
These fractional revivals  occur at times $t$ given by
$t \,=\, \pi \hbar r/ C_2\,s$
 where $r$ and $s$ are mutually prime integers.
It can be shown that at these instants the evolution operator $U$ can be
expressed as a finite sum of
operators $U_p$,
in each of which the phase factor multiplying the projection operator
$\kone{\phi_n} \, \langle \phi_n |$ is linear in $n$: That is, 
\begin{eqnarray}
U\left ( \pi \hbar r/ C_2\,s \right ) &=& 
\sum_{p=0}^{l-1} \, a_{p}^{(r,s)}
\;\;\; U_p \,,
\end{eqnarray}
with 
\begin{equation}
a_p^{(r,s)} \;=\; (1/l) \, \sum_{k=0}^{l-1} \, \exp \left [
- (i \, \pi \, k^2 r/ s) \, +\, (2\,i\, \pi \,
k p/l)
\right ]\;,
\end{equation}
and 
\begin{equation}
U_p \;=\;\sum_{n=0}^{\infty} \, \exp \,(-i \, n \theta_p) \;
\kone{\phi_n} \, \langle \phi_{n} |\;,
\;\;\theta_p \;=\;  \pi \,  \left[ (C_1 r/C_2 \, s) \, +\,
(2p/l) \right ] \;.
\end{equation}  
It is this decomposition of $U$ which is responsible for fractional
revivals, as
can be seen by investigating the action of the operator $U_p$ on the
initial state. For instance, if the  initial state \kone{\psi(0)} is
the ``coherent state" \kone{z} given by 
\begin{equation}
\kone{z}  \,=\, \exp \, \left ( -|z|^2/2
\right ) \;\; \sum_{n=0}^{\infty} \frac{z^n}{\sqrt{n!}} \;
\kone{\phi_n}\;,\; z \in
\BC\;,
\end{equation}
Eqs. (2.4)-(2.6) yield 
\begin{eqnarray}
\kone{\psi(\pi \hbar r/C_2 s)} &=&  \exp \, \left
(-i \pi C_0  r/C_2 \,s  \right ) \; \sum_{p=0}^{l-1}\;
a_p^{(r,s)}\; \; \kone{z \exp(-i \, \theta_p)}\,.
\end{eqnarray}
The state at time $\pi \hbar r / C_2 s$ is therefore a 
weighted sum of wave packets, each of which has the
same form as the initial state \kone{z}. 

It is clear from the foregoing that 
(i) fractional revivals arise
from the quadratic dependence of $E_n$ on $n$, and (ii) a  whole host of
such revivals of varying intensities can appear in a given case, depending
on the numerics, i.e., the precise values of $C_2$ and 
the integers $r$ and $s$ for which these revivals are detectable.
In the next two sections we show that, by  inducing suitable geometric
phases
in the basis states \kone{\phi_n}, we can eliminate these partial revivals
and restore near revivals at specific instants of time.

\section*{\large 3. SUPPRESSION OF FRACTIONAL REVIVALS}
\renewcommand{\theequation}{3.\arabic{equation}}
\setcounter{equation}{0}

We now consider the situation in which the Hamiltonian contains
a set of ``slow" parameters ${\bf R}$
that can be  varied adiabatically and cyclically with 
period $T$, as in Berry's original setting for the geometric
phase: i.e., ${\bf R}_T = {\bf R}_0$. 

In order to keep track of the parameter dependence in the basis states as
well, we denote by \kttwo{\phi_n}{\bf R} the $n^{\mbox{{\small th}}}$
eigenstate of
the
instantaneous Hamiltonian $H({\bf R})$. Then, at the end of each cycle of
period $T$, \kttwo{\phi_n}{\bf R} picks up a geometric phase $\gamma_n$
(over and above the $E_n$-dependent dynamical phase factor), given by the
expression$\,^{11)}$
\begin{eqnarray}
\gamma_n \;=\; i\, \oint \, \langle \phi_n \,, {\bf R} |
(\nabla_{\bf R} \kttwo{\phi_n}{\bf R}) \, \cdot \, d {\bf R}\;,
\end{eqnarray}
where the integral runs over the corresponding closed loop in parameter
space. Substituting for
\kttwo{\phi_n}{\bf R} at the end of a cycle, the time evolution operator
at time $T$ therefore becomes
\begin{equation}
U(T) \,=\,\sum_n  \exp \,\left [i \gamma_n \,- (i/\hbar) \, \int_0^T \,
dt \, E_n ({\bf R}_t) \, \right ] \, 
\kttwo{\phi_n}{{\bf R}_0} \, \langle \phi_n \,, {\bf R}_0|.
\end{equation}

To proceed,  we need to know the $n$-dependence of 
$\gamma_n$. For a general nonlinear spectrum $E_n$, one may expect a
dependence of the form
\begin{eqnarray}
\gamma_n &=& \Theta_0 \, +\, \Theta_1 \, n\,+\, \Theta_2 \, n^2\, + \,
\cdots
\end{eqnarray}
This  is generic, as it only  requires that $\gamma_n$ be a regular
function of $n$ (see Eqs. (3.13), (3.14) {\em ff.} below). As in the
expansion of
$E_n$, only terms up to $O(n^2)$
in Eq. (3.3) are relevant for our present purposes.
The coefficients $\Theta_i$ will clearly depend on
the manner in which the parameters ${\bf R}$ are varied.
We now substitute for
$\gamma_n$ and $E_n$ in Eq. (3.2) from Eqs. (3.3) and (2.2),
respectively, taking into account
  the fact that the
coeffcients $C_i$ are now time-dependent owing to the
variation of parameters in $H$.
Moreover, in each  cycle of period $T$
the state
\kttwo{\phi_n}{\bf R} picks up the same additional geometric and dynamical
phase.  Hence the time evolution operator at time $kT\,
(\mbox{where } k = 1, 2, \cdots)$ is given by
\begin{eqnarray}   
U(kT) &=&\sum_n  \exp \left [ i\, k (\nu_0 \,+\, \nu_1 n \,+ \,
\nu_2 n^2) \right ]
\;\kttwo{\phi_n}{{\bf R}_0} \, \langle \phi_n , {\bf R}_0|\,
\end{eqnarray} 
where
\begin{eqnarray}
\nu_i &= & \Theta_i
-(1/\hbar) \, \int_0^T C_i (t) \, dt \;, (i \,=\, 0,1,2)\;.
\end{eqnarray}     
From the discussion in Section 2, it is clear that all
fractional revivals of a  wave
packet will be eliminated if  the coefficient $\nu_2$ of $n^2$ in
the foregoing expression for  $U(kT)$
vanishes: this happens if  we arrange the 
variation of the parameters such that
\begin{eqnarray}
\Theta_2 &=& (1/\hbar) \int_0^T C_2(t) \, dt\,.
\end{eqnarray}
Once this is done, 
the exponent in $U(kT)$  has only terms that are linear in $n$.
Hence,  at
the relevant times $kT\;(k=1, 2, \cdots)$, the wave packet will no
longer
exhibit fractional revivals. 

To see in a little more detail how a geometric phase $\gamma_n$ of the
desired form  
may be
generated,$\,^{13)}$\, consider first 
the classical one-freedom Hamiltonian
\begin{equation}                                                                
H' (x',p') \;=\; A \, {p'}^2 \, +\, C \, V(x')\,                                
\end{equation} 
where $A$ and $C$ are positive constants, and $V(x')$ is a potential that
supports       
bounded, periodic motion. We assume that the quantum mechanical version,        
with a self-adjoint Hamiltonian $H'$, has a non-degenerate, discrete            
spectrum $\{E_n\}$ with normalized eigenstates $\{\kone{\chi_n}\}$. The         
position-space wavefunction $\langle x' \kone{\chi_n}\,=\, \chi_n (x')$         
can be chosen to be a real function, from which it follows that the Berry       
phase ${\gamma}_n$ that one may expect from the possible  variation of $A$
and $C$ vanishes identically.     
Reverting to the classical case, consider now the canonical transformation      
$(x' , p') \rightarrow (x,p)$ where                                             
\begin{equation}                                                                
x\;=\; x' \;, \;\; p\,=\, p' \, -\, (B/A) \, f(x')                        
\end{equation}                                                                  
where $B$ is a constant and the function $f(x)$ is to be specified. The         
transformed Hamiltonian is                                                      
\begin{equation}                                                                
H(x,p) \;=\; A \, p^2 \,+\,B\, [p\, f(x) \,+\, f(x) \, p] \, +\, C \, V(x)      
\, +\, (B^2/A) \, f^2 (x)\,,                                              
\end{equation}        
where we have written the cross terms in symmetric form in anticipation         
of quantization. Let $f(x)$ be chosen such that the phase trajectories          
corresponding to $H$ continue to represent bounded, periodic motion.            
Quantum mechanically, $H$ is 
obtained by the action upon $H'$ of the unitary operator     
\begin{equation}                                                                
W\;=\; \exp \, \left (-i B \hbar\, F(x) / A \right )\,,\;
F(x) \;=\; \int^{x} \, f(x)\,dx \,.                                             
\end{equation}              
Provided that this leads to a self-adjoint Hamiltonian $H$, it            
follows that $H$ and $H'$ are isospectral. The normalized eigenstates of        
$H$ are given by $\kone{\phi_n} \,=\, U\, \kone{\chi_n}$. Under an              
adiabatic, cyclic variation of parameters ${\bf R} \,=\, (A,B,C)$, the          
Berry phase acquired by \kone{\phi_n} is                                        
\begin{equation}                                                                
\gamma_n \;=\; i \, \oint \, \langle \phi_n |\, ( \nabla_{{\bf R}} \,           
\kone{\phi_n} ) \, \cdot \, d {\bf R}\,.                                        
\end{equation}         
As \kone{\phi_n} is normalized, this simplifies to 
\begin{eqnarray}                                                                
\gamma_n &=& i \, \oint  \langle \phi_n |\, ( \nabla_{{\bf R}}  U)
\kone{\chi_n}\,        
\cdot \, d {\bf R}\,.                                                           
\end{eqnarray}                                                                  
Using the unitarity of $U$, this yields                                         
\begin{eqnarray}                                                                
\gamma_n &=& (1/ \hbar)  \oint \, {\langle F \rangle}_n \,                 
\nabla_{{\bf R}}                                                                
(B/A) \, \cdot \, d {\bf R} \,,                                                 
\end{eqnarray}                                                                  
where                                                                           
\begin{eqnarray}                                                                
 {\langle F \rangle}_n &=& \int_{-\infty}^{\infty} \, F(x) \, \chi_n^2 (x)      
\, dx \, .                                                                      
\end{eqnarray}         
Therefore the $n$-dependence of the Berry phase $\gamma_n$ can be tailored
by choosing the transformation function $f(x)$ appropriately, even  
though $\{E_n\}$ remains the same for all acceptable choices of $f(x)$.

An explicit example is provided by the P\"{o}schl-Teller
Hamiltonian$\,^{14)}$\,
\begin{equation}
H' \;=\; Ap^2 - C \, {\mbox{sech}}^2 x\, \;\;\; (A,C > 0)
\end{equation}
which is unitarily transformed by
$W\, =\, \exp \,[-(i\, B/A \hbar) \ln \cosh x]$ to $H\,=\, W \,
H' \,
\ad{W}$ where 
\begin{equation}
H\;=\; A\, p^2 \, +\,B [ p \, (\tanh x) \, + \, (\tanh x) \, p]\,
- \,\left [ C + (B^2/A) \right ]  {\mbox{sech}}^2 x \,+\,
(B^2/A)\;.
\end{equation}
Writing $\eta \,=\, (1+4C/ A \hbar^2)^{1/2}$, the spectrum of $H$
is given by 
\begin{equation}
E_n \;=\;  -A \hbar^2 ( \eta-1 -2n)^2 /4
\end{equation}
where $n = 0,1, \cdots$, [$\frac{1}{2} (\eta -1)$] ([$\xi$] stands
for
the largest integer contained in  $\xi$).  As $E_n$ is quadratic
in $n$, a wave packet whose evolution is governed by $H$
will exhibit fractional revivals. Applying Eq. (2.4) to the case at
hand, we can now identify the instants at which such revivals occur as
 $ \pi\, r/A \hbar s$, where $r$ and $s$ are mutually prime integers.

To induce a geometric phase $\gamma_n$ in
the eigenstate \kone{\phi_n} of $H$,
it suffices in the present instance to vary just the two
parameters $A$ and $B$ 
with a  period $T$.
It can be shown$\,^{15)}$\, that
the geometric phase in this case   
is precisely of the form
\begin{equation}
\gamma_n \;=\; \Theta_o \,+\, \Theta_1 \,n \,+\, \Theta_2 \, n^2
\;.
\end{equation}
Putting in the exact expression for $\Theta_2$ in this instance, the
condition (3.6) for the cancellation of fractional revivals then
reads
\begin{eqnarray}
\int \, [ \, \nabla_{\bf R} (1/\eta^2) \, \times \,
\nabla_{\bf R} (B/A)\, ] \, \cdot \, d{\bf S} &=& \hbar^2  \int_0^T A(t)
\, dt\;, 
\end{eqnarray}
where the integral on the left runs over the surface bounded by the
closed loop in the $(A, B)$ plane along which $A$ and $B$ are varied.

Once fractional revivals are eliminated in the manner just described, we
are left with linear $n$-dependences for both $E_n$ and $\gamma_n$. But,
owing to the discretization of time in steps of $T$, this no longer
implies trivial periodicity, i.e., that full revivals would automatically
occur at integral multiples of $T$. However, as we shall show, {\em near}
revivals will now occur generically without any further fine tuning of the
parameters concerned.

\section*{\large 4. STATISTICS OF NEAR REVIVALS}
\renewcommand{\theequation}{4.\arabic{equation}}
\setcounter{equation}{0}

After the cancellation of the quadratic ($n^2$) term in the phase factors
in
Eq. (3.4), the effective time development operator at time $kT\; (k
=1,2,\cdots)$ is of the form 
\begin{eqnarray}
U(kT) &=& \exp (i k \nu_0) \, \sum_{n}  \exp (i k n \nu_1)
\;\kttwo{\phi_n}{{\bf R}_0} \, \langle \phi_n , {\bf R}_0|\,.
\end{eqnarray}
Therefore an initial coherent state $\kone{\psi(0)}\,=\, \kone{z}$ 
given by Eq. (2.7) evolves to
\begin{eqnarray}
\kone{\psi(kT)} &=& \kone{z \, \exp (i k \nu_1)}\;.
\end{eqnarray} 
Correspondingly, the correlation function given by Eq. (1.1) becomes
\begin{eqnarray}
{\cal C}(kT) &=& \exp\left [\,
2|z|^2\, (\cos k \nu_1 -1)\,\right ]\;.
\end{eqnarray}
Thus, if $k\nu_1$ happens to be an integer multiple of $2 \pi$
(recall that $\nu_1$ depends on $T$, {\em cf.} Eq. (3.5)), then full
revivals occur with $T_{rev} \,=\, kT$ as the basic revival time. In
general, however, $\nu_1$ is an irrational number (modulo $2 \pi$).
Therefore, if $\theta_0$ is the phase of the complex number $z$ labelling
the initial state \kone{\psi(0)}, and $\theta_k$ that of the corresponding
number at time $kT$, Eq. (4.2) shows that the (discrete time) evolution of
the state is entirely equivalent to the circle map 
\begin{eqnarray}
\theta_{k} &=& \theta_{k - 1} \,+\, \nu_1   \;\;\; \mbox{(mod 2$\pi$)}
\end{eqnarray}
corresponding to a rigid ``irrational" rotation.
As is well known,$\,^{16)}$\, this map has no periodic orbits,
is ergodic, and has a uniform invariant density. Given an
$\epsilon$-neighborhood $I_{\epsilon}$ of the initial phase $\theta_0$, 
we have a
Poincar\'{e}  recurrence to $I_{\epsilon}$ at time $kT$ if $\theta_k \in
I_{\epsilon}$. This corresponds to a near revival of the original wave
packet, because it is easily shown that we then have
\begin{equation}
{\cal C}(kt) \, > \, 1 - |z|^2 \, \epsilon^2 \;.
\end{equation}
\mbox{The\,\,\,\,recurrence\,\,\,\,\,statistics\,\,for\,\,the\,\,
rotation\,\,\, 
map\,\,is\,
already\,\,\,known\,\,\,\,from\,\,
certain ``gap}\\
theorems",$\,^{17)}$\, and has found applications in the
study
of level spacings of oscillator systems$\,^{18)}$\, and 
recurrences in the coarse grained dynamics of quasiperiodic flow on a
two-torus.$\,^{19)}$\, In earlier work,$\,^{20)}$\, we have discussed the
application of these theorems to near revivals of coherent states of a
deformed (displaced, squeezed) oscillator, and these results can now be
taken over directly to the problem at hand. While the mean time 
between successive near revivals turns out to be simply $2 \pi T/
\epsilon$ 
in accordance with the ergodic theorem (the measure of $I_{\epsilon}$
being $\epsilon/2 \pi$), the probability distribution of 
this time interval   is quite remarkable. It is typically
concentrated at three specific values $k_1 T,\,k_2 T$ and $(k_1+k_2)T$,
regardless of the values of $\nu_1$ and $\epsilon$. The actual values of
the integers $k_1$ and $k_2$, and the relative frequencies of occurrence
of the three different revival times, are of course dependent on 
$\nu_1$ and $\epsilon$.$\,^{20)}$

We have thus established an interesting link between revivals,
anholonomies and recurrences. These are manifestations, respectively, of
quantum interference efffects, non-trivial topology in parameter space,
and ergodic behavior in a discrete-time classical dynamical system. The
distinctive nature of the generic distribution of near revival times that
we have predicted should provide a clear signature, from the experimental
point of view, for testing the validity of our formulation of certain
aspects of wave packet evolution.

\section*{\large ACKNOWLEDGMENT}
SL thanks the Department of Science and Technology, India, for partial 
support under the grant SP/S2/E-03/96. 

\end{document}